\documentclass[10pt,twocolumn,a4paper,secnumarabic]{revtex4}
\usepackage{graphicx}

\begin{document}
\bibliographystyle{wp3}
  \title{Relational Grid Monitoring Architecture (R-GMA)}

  \author{Andrew Cooke}
  \author{Werner Nutt}
  \affiliation{Heriot-Watt, Edinburgh, UK}
                                                                                
  \author{James Magowan}
  \author{Paul Taylor}
  \affiliation{IBM UK Ltd.}

  \author{Jason Leake}
  \affiliation{Objective Engineering Ltd.}
                                                                                
  \author{Rob Byrom}
  \author{Laurence Field}
  \author{Steve Hicks}
  \author{Manish Soni}
  \author{Antony Wilson}
  \affiliation{ PPARC, UK}
                                                                                
  \author{Roney Cordenonsi}
  \affiliation{Queen Mary, University of London, UK}
                                                                                
  \author{Linda Cornwall}
  \author{Abdeslem Djaoui}
  \author{Steve Fisher}
  \email[email: ]{s.m.fisher@rl.ac.uk} 
  \affiliation{Rutherford Appleton Laboratory, UK}
                                                                                
  \author{Norbert Podhorszki}
  \affiliation{SZTAKI, Hungary}
                                                                                
  \author{Brian Coghlan}
  \author{Stuart Kenny}
  \author{David O'Callaghan}
  \author{John Ryan}
  \affiliation{Trinity College Dublin, Ireland}

  \begin{abstract}
    We describe R-GMA (Relational Grid Monitoring Architecture) which
    has been developed within the European DataGrid Project as a Grid
    Information and Monitoring System. Is is based on the GMA from
    GGF, which is a simple Consumer-Producer model. The special
    strength of this implementation comes from the power of the
    relational model. We offer a global view of the information as if
    each Virtual Organisation had one large relational database. We
    provide a number of different Producer types with different
    characteristics; for example some support streaming of
    information. We also provide combined Consumer/Producers, which
    are able to combine information and republish it. At the heart of
    the system is the mediator, which for any query is able to find
    and connect to the best Producers for the job.  We have developed
    components to allow a measure of inter-working between MDS and
    R-GMA.  We have used it both for information about the grid
    (primarily to find out about what services are available at any
    one time) and for application monitoring. R-GMA has been deployed
    in various testbeds; we describe some preliminary results and
    experiences of this deployment.
  \end{abstract}
  \maketitle

  \section{Introduction}

\begin{figure}[htfb]
    \includegraphics[scale=0.4]{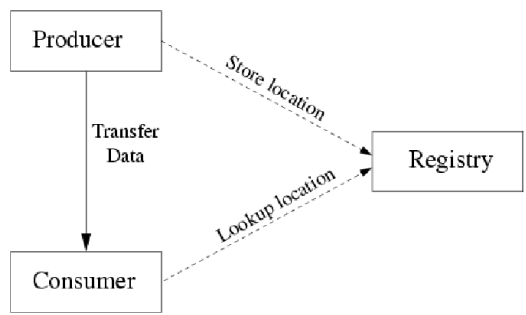}
    \caption{Grid Monitoring Architecture}
    \label{figure-gma}
  \end{figure}

  The Grid Monitoring Architecture (GMA)\cite{ref:perf-arch} of the
  GGF, as shown in Figure~\ref{figure-gma}, consists of three
  components: \textit{Consumers}, \textit{Producers} and a directory
  service, which we prefer to call a \textit{Registry}).

  In the GMA Producers register themselves with the Registry and
  describe the type and structure of information they want to make
  available to the Grid. Consumers can query the Registry to find out
  what type of information is available and locate Producers that
  provide such information. Once this information is known the
  Consumer can contact the Producer directly to obtain the relevant
  data.  By specifying the Consumer/Producer protocol and the
  interfaces to the Registry one can build inter-operable
  services. The Registry communication is shown on
  Figure~\ref{figure-gma} by a dotted line and the main flow of data
  by a solid line.

  The current GMA definition also describes the registration of
  Consumers, so that a Producer can find a Consumer. The main reason
  to register the existence of Consumers is so that the Registry can
  notify them about changes in the set of Producers that interests
  them.

  The GMA architecture was devised for monitoring but we think it
  makes an excellent basis for a \emph{combined} information and
  monitoring system. We have argued before\cite{ref:time-bncod-2001}
  that the only thing which characterises monitoring information is a
  time stamp, so we insist upon a time stamp on all measurements -
  saying that this is the time when the measurement was made, or
  equivalently the time when the statement represented by the tuple
  was true.

  The GMA does not constrain any of the protocols nor the underlying
  data model, so we were free when producing our implementation to
  adopt a data model which would allow the formulation of powerful
  queries over the data.

  R-GMA is a relational implementation of the GMA, developed within the
  European DataGrid (EDG), which brings the power and flexibility of
  the relational model. R-GMA creates the impression that you have one
  RDBMS per Virtual Organisation (VO). However it is important to
  appreciate that what our system provides, is a way of using the
  relational model in a Grid environment and that we have \emph{not}
  produced a general distributed RDBMS. All the producers of
  information are quite independent. It is relational in the sense
  that Producers announce what they have to publish via an SQL CREATE
  TABLE statement and publish with an SQL INSERT and that Consumers
  use an SQL SELECT to collect the information they need. For a more
  formal description of R-GMA see the forthcoming CoopIS
  paper\cite{ref:rgma-coopis-2003}.

  R-GMA is built using servlet technology and is being migrated
  rapidly to web services -- specifically to fit into an
  OGSA\cite{ref:ogsa-spec} framework.

  \section{Query types and Producer Types}
  
  We have so far defined not just a single Producer but five different
  types: a DataBaseProducer, a StreamProducer, a ResilientProducer, a
  LatestProducer and a CanonicalProducer. All appear to be Producers
  as seen by a Consumer - but they have different characteristics.
  The CanonicalProducer, though in some respects the most general, is
  somewhat different as there is no user interface to publish data via
  an SQL \texttt{INSERT} statement. Instead it triggers user code to
  answer an SQL query. The other Producers are all
  \texttt{Insertable}; this means that they all have an interface
  accepting an SQL \texttt{INSERT} statement.

  The other producers are instantiated and given the description of
  the information they have to offer by an SQL \texttt{CREATE TABLE}
  statement and a \texttt{WHERE} clause expressing a predicate that is
  true for the table. Currently this is of the form \texttt{WHERE
  (column\_1=value\_1 AND column\_2=value\_2 AND ...)}.  To publish
  data, a method is invoked which takes the form of a normal SQL
  \texttt{INSERT} statement.

  Three kinds of query are supported: History, Latest and
  Continuous. The history query might be seen as the more traditional
  one, where you want to make a query over some time period -
  including ``all time''. The latest query is used to find the current
  value of something and a continuous query provides the client with
  all results matching the query as they are published. A continuous
  query is therefore acting as a filter on a published stream of data.

  The DataBaseProducer supports history queries. It writes each record
  to an RDBMS. This is slow (compared to a StreamProducer) but it can
  handle joins.  The StreamProducer supports continuous queries and
  writes information to a memory structure where it can be picked up
  by a Consumer. The ResilientStreamProducer is similar to the
  StreamProducer but information is backed up to disk so that no
  information is lost in the event of a system crash. The
  LatestProducer supports latest queries by holding only the latest
  records in an RDBMS.

  Each record has a time stamp, one or more fields which define what
  is being measured (e.g. a hostname) and one or more fields which are
  the measurement (e.g. the 1 minute CPU load average). The time stamp
  and the defining fields are close to being a primary key - but as
  there is no way of knowing who is publishing what across the Grid,
  the concept of primary key (as something globally unique) makes no
  sense. The LatestProducer will replace an earlier record having the
  same defining fields, as long as the time stamp on the new record is
  more recent, or the same as the old one.

  Producers, especially those using an RDBMS, may need cleaning from
  time to time.  We provide a mechanism to specify those records of a
  table to delete by means of a user specified SQL \texttt{WHERE}
  clause which is executed at intervals which are also specified by
  the user.  For example it might delete records more than a week old
  from some table or it may only hold the newest one hundred rows, or
  it might just keep one record from each day.

  \begin{figure*}[htfb]
    \includegraphics[scale=0.405]{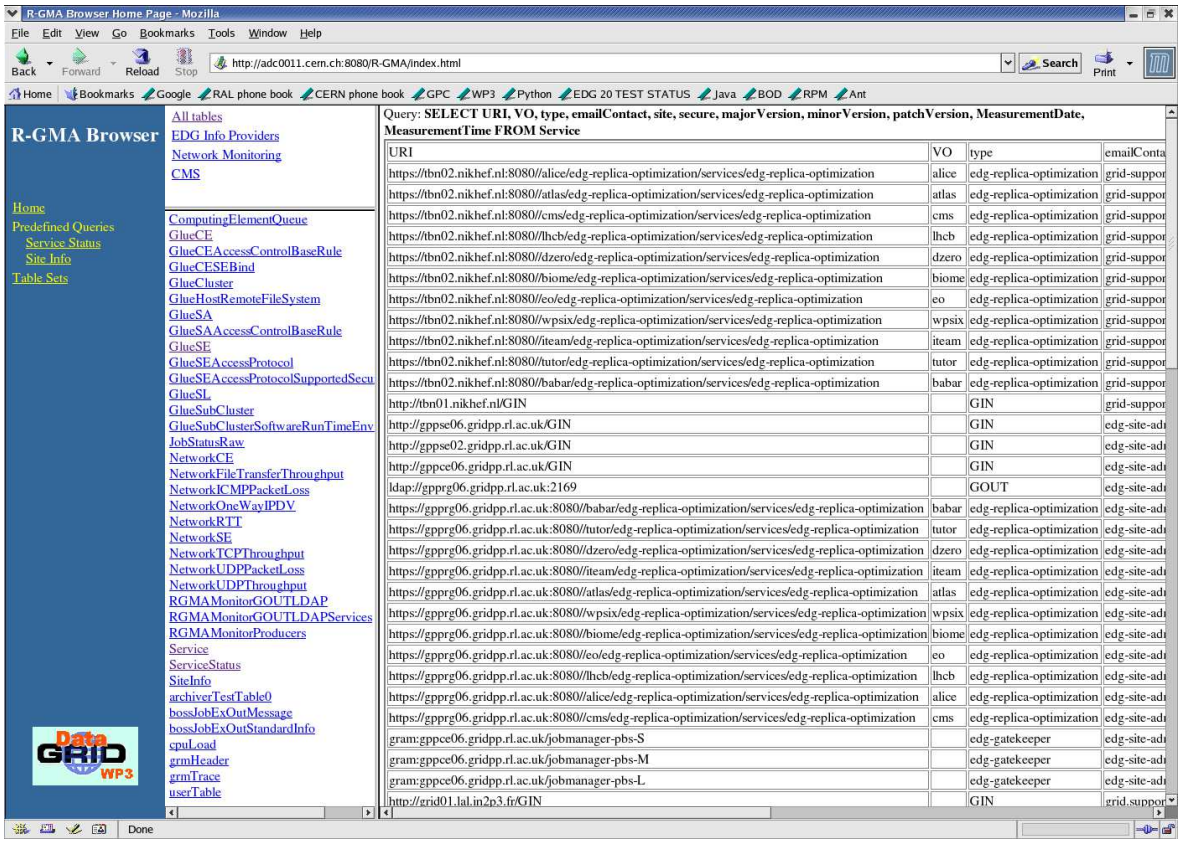}
    \caption{R-GMA BrowserServlet}
    \label{figure-browser}
  \end{figure*}

\begin{figure}[htfb]
    \includegraphics[scale=0.5]{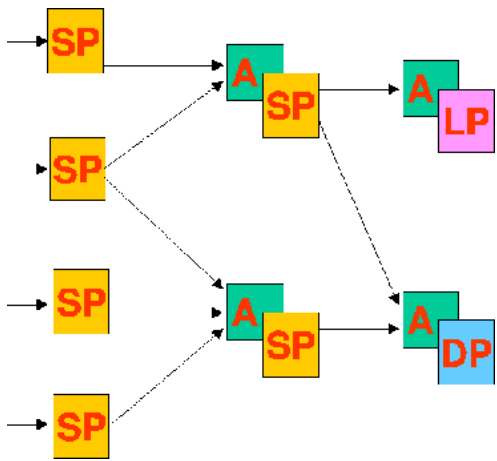}
    \caption{A possible topology of R-GMA components}
    \label{figure-topo}
  \end{figure}

  Another valuable component is the Archiver which is a combined
  Consumer-Producer.  You just have to tell an Archiver what to
  collect and it does so on your behalf.  An Archiver works by taking
  over control of an existing Producer and instantiating a Consumer
  for each table it is asked to archive. This Consumer then connects
  via the mediator to all suitable Producers and data starts streaming
  from those Producers, through the Archiver and into the new
  Producer. The inputs to an Archiver are always streams from a
  StreamProducer or a ResilientStreamProducer.  It will re-publish to
  any kind of \texttt{Insertable}. This allows useful topologies of
  components to be constructed such as the one shown in
  Figure~\ref{figure-topo}

  This shows a number of StreamProducers (labelled SP) which is
  normally the entry point to R-GMA. There is then a layer of
  Archivers (A) publishing to another StreamProducer. Finally there is
  an Archiver to a LatestProducer (LP) and an Archiver to a
  DataBaseProducer (DP) to answer both Latest and History queries.

  We intend to allow some kinds of producer to answer more than one
  kind of query - but for now we are keeping it simple.

  \section{Tools}
  There are a number of tools available to query R-GMA Producers.
  There is a command line tool, a Java graphical display tool, and the
  R-GMA Browser. The browser is accessible from a Web browser without
  any R-GMA installation. It offers a few custom queries, and makes it
  easy for you to write your own. A screen shot is shown in
  Fig~\ref{figure-browser}.

  The command line tool, which is written in Python, is the most
  powerful.  It is designed to do simple things very easily - but if
  you want to carry out more complex operations you must code them
  yourself using one of the APIs. It supports one instance of each
  kind of producer and one Archiver at any one time. You can also find
  what tables exist, find details of a table and issue any kind of
  query.

  \begin{figure*}[htfb]
      \includegraphics[scale=0.56]{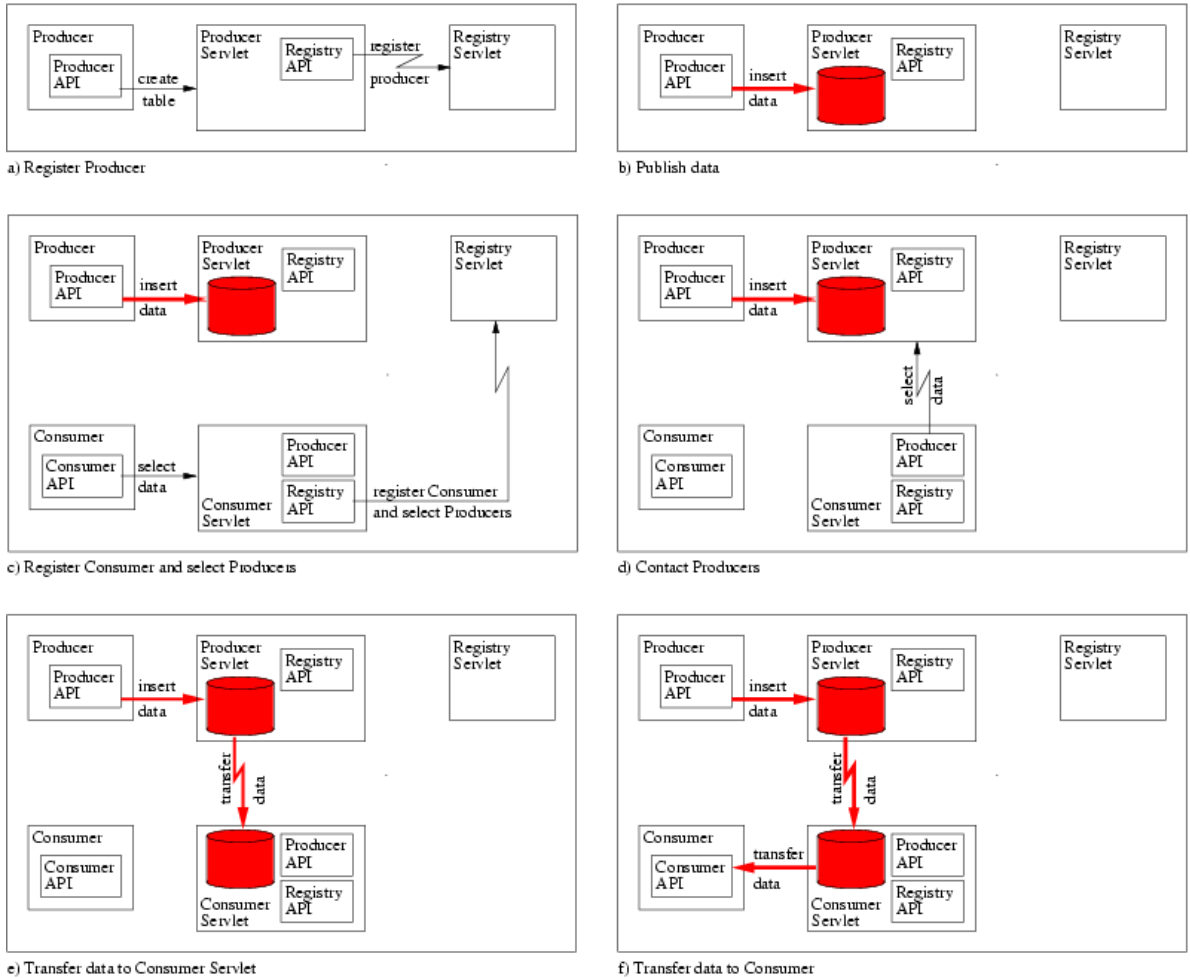}
      \caption{Relational Grid Monitoring Architecture}
    \label{figure-rgma}
  \end{figure*}

  \section{The registry and the mediator}
  The registry stores information about all producers currently
  available. Currently there is only one physical Registry per
  VO. This bottleneck and single point of failure is being
  eliminated. Code has been written to allow multiple copies of the
  registry to be maintained. Each one acts as master of the
  information which was originally stored in that Registry instance
  and has copies of the information from other Registry
  instances. Synchronisation is carried out frequently. Currently VOs
  are disjoint, we plan to allow information to be published to a set
  of VOs.

  The mediator (which is hidden behind the Consumer interface) is the
  component which makes R-GMA easy to use.  Producers are associated with
  views on a virtual data base. Currently views have the form:

  \begin{quote}
    SELECT * FROM $<$table$>$ WHERE $<$predicate$>$
  \end{quote}

  This view definition is stored in the Registry.  When queries are
  posed, the Mediator uses the Registry to find the right Producers and
  then combines information from them.

  \section{Architecture}
  R-GMA is currently based on Servlet technology. Each component has
  the bulk of its implementation in a Servlet.  Multiple APIs in Java,
  C++, C, Python and Perl are available to communicate with the
  servlets. The basic ones are the Java and C++ APIs which are
  completely written by hand. The C API calls the C++ and the Python
  and Perl are generated by SWIG. We make use of the Tomcat Servlet
  container. Most of the code is written in Java and is therefore
  highly portable. The only dependency on other EDG software
  components is in the security area.

  Figure~\ref{figure-rgma} shows the communication between the APIs
  and the Servlets.  When a Producer is created its registration
  details are sent via the Producer Servlet to the Registry
  (Figure~\ref{figure-rgma}a). The Registry records details about the
  Producer, which include the description and view of the data
  published, \emph{but not the data itself}.  The description of the
  data is actually stored as a reference to a table in the Schema. In
  practise the Schema is co-located with the Registry.
  Then when the Producer publishes data, the data are transferred to a
  local Producer Servlet (Figure~\ref{figure-rgma}b).

  When a Consumer is created its registration details are also sent to
  the Registry although this time via a Consumer Servlet
  (Figure~\ref{figure-rgma}c). The Registry records details about the
  type of data that the Consumer is interested in.  The Registry then
  returns a list of Producers back to the Consumer Servlet that match
  the Consumers selection criteria.

  The Consumer Servlet then contacts the relevant Producer Servlets to
  initiate transfer of data from the Producer Servlets to the Consumer
  Servlet as shown in \mbox{Figures~\ref{figure-rgma}d-e}.

  The data are then available to the Consumer on the Consumer Servlet,
  which should be close in terms of the network to the Consumer
  (Figure~\ref{figure-rgma}f).

  As details of the Consumers and their selection criteria are stored
  in the Registry, the Consumer Servlets are automatically notified
  when new Producers are registered that meet their selection
  criteria.

  The system makes use of soft state registration to make it
  robust. Producers and Consumers both commit to communicate with
  their servlet within a certain time. A time stamp is stored in the
  Registry, and if nothing is heard by that time, the Producer or
  Consumer is unregistered. The Producer and Consumer servlets keep
  track of the last time they heard from their client, and ensure that
  the Registry time stamp is updated in good time.

  \section{Applications of R-GMA}
  R-GMA has applications right across the Grid.

  For example it is being used for network monitoring where the
  flexibility of the relational model offers a more natural
  description of the problem. The results of the monitoring are being
  used to compute the relative costs (in time) of moving data between
  two points within DataGrid to optimise use of resources.

  CMS\cite{ref:CMS-www}, one of the forthcoming experiments at CERN
  has identified the need to monitor the large numbers of jobs that
  are being executed simultaneously at multiple remote sites. They
  have adapted their BOSS job submission and tracking system which
  previously wrote to a well known RDBMS to simply publish the job
  status information via R-GMA\cite{ref:boss-ieee-2003}.

  Some other applications are explained below.
 
  \subsection{MDS replacement}
  First it can be used as a replacement for MDS\cite{ref:mds-www}. A
  small tool (GIN) has been written to invoke the MDS-like EDG
  info-providers and publish the information via R-GMA. The
  info-provider is a small script which can be invoked to produce
  information in LDIF format. All our information providers conform to
  the GLUE schemas\cite{ref:glue-www} Another tool (GOUT) is available
  to republish R-GMA data to an LDAP server for the benefit of legacy
  applications. However we expect that most applications will wish to
  benefit from the power of relational queries. GOUT is an Archiver
  with a Consumer which periodically publishes to an LDAP
  database. Both GIN and GOUT are driven by configuration files which
  define the mapping between the LDAP schema and the relational
  schema.
 
  \subsection{Service location and monitoring}
  We has defined a pair of tables: Service and ServiceStatus. This is
  a rather common pattern where some rapidly changing attributes have
  been separated off into a separate status table. In this case the
  person responsible for the provision of the service publishes its
  existence and how to contact it into the Service table. Each Service
  tuple includes the type of the service and a URI for the service
  where the hostname within the URI is where the serice is
  located. (Eventually these will all be URLs to contact the service)

  Each service provider specifies a command (as a function of the
  service type) which can be run to obtain the ServiceStatus. This
  is invoked locally on each machine running a service. The
  information is then collected by an Archiver to a LatestProducer. So
  the Service table says what should exist and the ServiceStatus gives
  the current state Grid wide.

  Finally we use Nagios\cite{ref:nagios-www}, an open source host,
  service and network monitoring program, to display graphs showing
  the reliability of the various services. Nagios reconfigures itself
  periodically to look at the information provided by the known
  Services in the Service table and collects information on the Status
  by looking at the ServiceStatus information. Nagios is then able to
  issue warnings to sysadmins as appropriate. This is completely table
  driven using the information in these two tables.

  \subsection{Application monitoring of parallel applications}
  GRM\cite{ref:grm-sanfrancisco-2001} is an on-line monitoring tool
  for parallel applications executed in the grid environment (or in a
  cluster, or on a supercomputer). PROVE is an on-line trace
  visualisation tool for parallel/distributed message-passing
  applications executed in the grid environment. It processes trace
  data generated by GRM.

  The Mercury monitor\cite{ref:mercury-klagenfurt-2003} is the
  monitoring system developed within the Gridlab project. The
  gridified version of GRM uses Mercury to transfer the large amount
  of trace data from the execution machines to the user's machine.
  Mercury currently consists of local monitor (LM) services running on
  each execution machine and a main monitor service (MM) on the
  front-end-node of a cluster/supercomputer. Different
  clusters/supercomputers in the grid have their own independent
  Mercury installation and they work independently from each other.
  
  When the application (instrumented with GRM calls) is submitted to
  the grid, the site for execution is chosen by a resource broker. The
  user (and GRM) does not know the site in advance. When the
  application is started, it registers in Mercury but GRM does not
  know where to connect, i.e. the address of the corresponding main
  monitor service running on the execution site.
  
\begin{figure}[htfb]
    \includegraphics[scale=0.33]{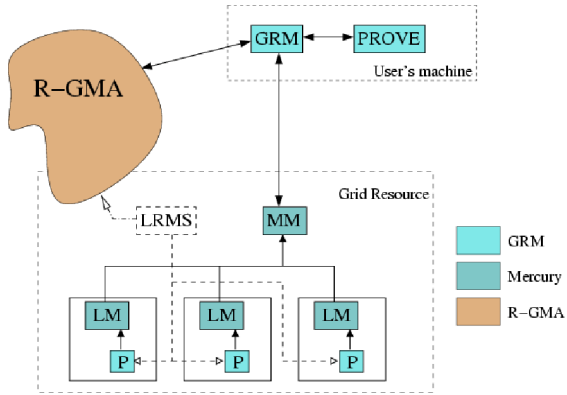}
    \caption{GRM, Mercury and R-GMA}
    \label{figure-grm-mercury-rgma}
  \end{figure}

  To solve this problem, R-GMA is used as shown in
  Fig.~\ref{figure-grm-mercury-rgma}.  Applications are
  registered in R-GMA with their global job ID by the local resource
  management system (LRMS) and the corresponding Mercury monitor
  address, just before they are launched. GRM looks for the user's
  application in R-GMA based on the global job ID. When it is found,
  the monitor address is used to establish the connection between GRM
  and Mercury. After that, streaming of trace data through Mercury can
  be started.

  \section{Results so far}
  Unfortunately we have few results to offer at this stage. It has
  taken some time to get from the state of having something which
  passes all its unit tests (about 400 for the Java API) to a stable
  distributed system - which we think we now have. We have recently
  started running performance tests to understand the behaviour of the
  code.  We have so far tested with many StreamProducers, and one
  Archiver feeding into a LatestProducer which is then queried to make
  sure that the Archiver is keeping up with the total flow of
  data. This showed up a few bottlenecks, but the biggest one was the
  I/O. To avoid this problem, new code is being developed to make use
  of the new \texttt{java.nio} package which offers non-blocking
  I/O. With this in place early measurements indicate that with Producers
  publishing data following the pattern expected of a ``typical'' site
  having an SE (Storage Element) and 3 CEs (Computing Elements) we
  will be able to support around 150 sites with this simple topology.

  To achieve better performance we may need a layer of Archivers
  combining streams into bigger streams so as to limit the fan-in to
  any one node. The other way to obtain significantly better
  performance is not to attempt to get all the information into one
  place. As the mediator becomes more powerful, it will be able to
  make use of multiple LatestProducer archives, and carry out a
  distributed query over them. We hope to benefit from developments in
  OGSA-DAI\cite{ref:ogsa-dai-www} in this area.

  For testing our performance in a testbed we use both a ``private''
  R-GMA testbed which is distributed over multiple sites and the main
  EDG development testbed. We try to test our software on the private
  testbed before passing it on. Consequently both testbeds are highly
  unstable: sites come and go and software is continuously updated. So
  the challenge is to make meaningful measurements on an ever changing
  system. Our approach is to monitor the Computing and Storage
  elements information by observing all the intermediate
  components. The mechanism does not rely upon configuration files
  giving all the expected components. Information on response times
  and availability and age of information at various points in the
  system is collected and published to a DataBaseProducer. Another
  program is being developed to try and make sense of this information
  and produce information each hour for the previous 24 hours.  These
  results will in turn be published and probably fed into Nagios to
  help identify any trends graphically.

  The effort involved in making meaningful measurements on such a
  system as R-GMA should not be underestimated!

  \section{Future of R-GMA}
  RGMA currently uses Servlet Technology for its underlying
  implementation.  This means for example that a Producer servlet
  keeps track of the many Producers instances that may actually be
  running within this container.  Developments over the last 1-2 years
  have highlighted the advancement and uptake of web services, indeed
  GGF has supported investigations and a proposed Specification (OGSI)
  looking into Grid Services. This effectively takes Grid requirements
  and concepts and specifies how web services can be used to achieve
  these requirements. 
                                                                                
  The Open Grid Services Architecture (OGSA) was proposed within the
  GGF for developing a Grid environment based upon Web Services and
  this has gradually received acceptance within the Grid Community.

  OGSI builds on top of web services standards and defines a 'Grid
  service' as Web services that must implement a mandatory interface
  (GridService) and may implement additional ones.  Grid services that
  conform to the OGSI specification can be invoked by any client or
  any other Grid service that follows the conventions, subject to
  policy and compatible protocol bindings.  Now that OGSI is maturing
  with version 1.0 of the specification nearing its final release, we
  feel the time is right to start moving in this direction.

  To this end we are starting to move our schema and registry towards
  Web Services which will work within an OGSA environment.
                                                                                
  Using OGSI factories for creating instances instead of servlets
  provide easier lifetime management, identity tracking and state
  management.  Initially the interfaces for R-GMA Grid services are
  wrapping the classes used within the existing servlets, so as to
  maintain backward compatibility and evolve the two versions in
  parallel.
 
  \section{Conclusion}
  We have a useful architecture and an effective implementation with a
  number of components which work well together. We expect that R-GMA
  will have a long, happy and useful life, both in its current form
  and when reincarnated within an OGSA framework. For more details of
  R-GMA, please see: \texttt{http://hepunx.rl.ac.uk/edg/wp3/} or in the near
  future: \texttt{http://www.r-gma.org/}.

  \begin{acknowledgments}
    We wish to thank our patient users, the EU and our national
    funding agencies.
  \end{acknowledgments}

  \bibliography{rgma}

\end{document}